\documentclass[showpacs,aps,pra,amsmath,amssymb,twocolumn]{revtex4-1}
\usepackage{graphicx}
\usepackage{dcolumn}
\usepackage{bm,color}
\usepackage{txfonts}
\usepackage{amsmath,epsfig}
\usepackage{epstopdf}

\newcommand{\eq}{Eq.~}
\newcommand{\eqs}{Eqs.~}
\newcommand{\fig}{Fig.~}

\newcommand{\cf} {cf.~}
\newcommand{\ug} {\!=\!}

\newcommand{\piu} {\!+\!}
\newcommand{\meno} {\!-\!}
\newcommand{\ie} {i.e.~}
\newcommand{\eg} {e.g.~}
\newcommand{\rhs} {r.h.s.~}

\newcommand{\rref} {Ref.~}
\newcommand{\rrefs} {Refs.~}

\newcommand{\ibid}{{\it ibid.~}}  
\newcommand{\cfr} {cfr.~}

\begin{document}

\pacs{03.67.Lx, 03.67.Hk, 42.50.-p}

\title{Quasideterministic realization of a universal quantum gate {in} a single scattering process}

\author{F. Ciccarello\mbox{$^{1,3}$}}
\author{D. E. Browne\mbox{$^{2}$}}
\author{L. C. Kwek\mbox{$^{3}$}}
\author{H. Schomerus\mbox{$^{4}$}}
\author{M. Zarcone\mbox{$^{5}$}}
\author{S. Bose\mbox{$^{2}$}}
\affiliation{\mbox{$^{1}$}Scuola Normale Superiore, Piazza dei Cavalieri, 7, I-56126 Pisa, Italy\\
\mbox{$^{2}$}Department of Physics and Astronomy, University College London, Gower Street, London WC1E 6BT, United Kingdom \\
\mbox{$^{3}$}Centre for Quantum Technologies, National University of Singapore, 3 Science Drive 2, Singapore 117543 \\
\mbox{$^{4}$}Department of Physics, Lancaster University, Lancaster, LA1 4YB, United Kingdom \\
\mbox{$^{5}$} CNISM and Dipartimento di Fisica, Universit\`{a} degli Studi di Palermo, Viale delle Scienze, Ed.~18, 
I-90128 Palermo, Italy}
\begin{abstract}
We show that a flying particle, such as an electron or a photon, scattering along a one-dimensional waveguide from a pair of static spin-1/2 centers, such as quantum dots, can implement a CZ gate (universal for quantum computation) between them. This occurs quasi-deterministically in a single scattering event, hence with no need for any post-selection or iteration, {and} without demanding the flying particle to bear any internal spin. We show that an easily matched hard-wall boundary condition along with the elastic nature of the process are key to such performances.
\end{abstract}

\maketitle
\noindent

Interfacing static qubits \cite{nc} mediated by flying particles is a prominent paradigm  in the quest for efficient ways to implement quantum information processing (QIP) \cite{distributed, rus}. 
As a major motivation, this is the only way to {\it jointly} address quantum registers located {\it far} from each other, {thus} featuring no direct mutual interaction ({this} is usually sought to favor local addressing). Within this general framework, over the past few years a research line has thrived around the idea that the crosstalk between the static qubits can be mediated by particles {\it scattering} from them \cite{imps1,imps2,imps3,imps4,imps5,imps6,imps7,imps8,imps9, lasphys, intj, burgarth}.

Yet, all of such strategies unavoidably face an inherent {major} drawback. For a given quantum task  to be efficiently accomplished, the link between the static objects should occur by means of the local interaction of each static object with a {\it quantum} flying bus. Namely, this should feature inherently quantum motional and/or internal (pseudo) spin degrees of freedom (DsOF). To do so, however, the coupling between the flying and static particles will {in general} entangle them so as to bring about decoherence affecting the DsOF of the static objects. Owing to such effect, the attainment of satisfactory figures of merit thus demands for further actions to complement the above interaction. These typically comprise iterated injection of the flying particles and post-measurements over their DsOF \cite{distributed, rus, imps1,imps2,imps3,imps4,imps5,imps6,imps7,imps8,imps9, lasphys, intj, burgarth}. While such {bus}-dynamics conditioning generally enhances the performances, it usually comes at the cost of making the process probabilistic and may be demanding in practice (\eg spin post-selection of mobile-electrons in semiconducting media \cite{loss}). 

Moreover, as far as {\it scattering} scenarios are concerned, there appear intrinsic hindrances to accomplish certain tasks such as the implementation of a two-qubit quantum gate (TQG), typically the most challenging {and essential} process in most QIP architectures especially when allowing for universal quantum computation (QC). To see this, assume that we need to realize a TQG between a flying qubit and a scattering center (SC) endowed with spin in a one-dimensional (1D) waveguide \cite{guillermo}. 
Even if the dynamics is conditioned to either the reflection or transmission channel, the resulting process lacks unitarity, \ie a paramount prerequisite for a TQG, unless quite specific regimes of parameters and, importantly, interaction models are addressed \cite{guillermo}. Analogous considerations {\it a fortiori} hold when many scattering centers are present and a gate involving their DsOF is sought, which will be our focus in this work. {Scattering-based scenarios thus appear as adverse arenas to perform quantum {\it algorithms}, which arguably explains why mere entanglement generation was almost exclusively investigated to date \cite{imps1,imps2,imps3,imps4,imps5,imps6,imps7,imps8,imps9, lasphys,intj,burgarth}}.
Nevertheless, scattering-based implementations are attractive {because} of the low demand for control. One normally just needs to set the itinerant-bus wave vector and wait for the collision to occur, thus bypassing any interaction-time tuning ({usually a significant noise source}). Further benefits such as the resilience to relevant detrimental factors including static disorder \cite{lasphys}, phase noise \cite{intj} and imperfect particle-wave-vector setting \cite{burgarth} have been shown. {Except for the attempt in \rref\cite{guillermo} such advantages have so far been harnessed solely for mere entanglement generation \cite{imps1,imps2,imps3,imps4,imps5,imps6,imps7,imps8,imps9, lasphys,intj,burgarth}.

{Here, we discover a simple strategy for the realization of quantum gates between static qubits through a particle scattering from them. The injection of the latter, which is not demanded to bear any internal DOF, followed by its multiple scattering from the SCs suffice to {quasi-\it deterministically} achieve the gate in one shot. Also, {\it neither post-selection of any kind nor repeated sending of the flying mediators} are required. Thus, besides unveiling an unexpected suitability of scattering-based methods \cite{imps1,imps2,imps3,imps4,imps5,imps6,imps7,imps8,imps9, lasphys,intj,burgarth} to achieve unitary operations, our work sets a milestone within the distributed-QIP context \cite{distributed, rus}. For two static qubits a {\it universal} CZ gate naturally arises showing the effectiveness of our scheme.}

{\it Central idea.} 
Consider a monochromatic spinless particle $f$ of wave vector $k$ propagating along a 1D wire that impinges on an array of SCs [see \fig1(a)]. Once multiple scattering has occurred and assuming, importantly, that the process is elastic, $f$ can only be found either reflected or transmitted with wave vectors $\!-\!k$ and $k$, respectively [see \fig2(b)]. Let $\{|\nu\rangle\}$  be a basis of the SCs'~Hilbert space and $|\mu\rangle$ one of its elements, whereas $|{\pm k}\rangle$ are momentum eigenstates of $f$. Let $|\Psi_{in}\rangle\ug|k\rangle|\mu\rangle$ be the overall system's initial state. As $f$ is scattered off the final state reads
\begin{equation}\label{psif}
|\Psi_f\rangle\ug|\meno k\rangle\sum_\nu r_{\nu\mu}|\nu\rangle+|k\rangle\sum_\nu t_{\nu\mu}|\nu\rangle\,\,,
\end{equation}
where $r_{\nu\mu}$ ($t_{\nu\mu}$) is a reflection (transmission) probability amplitude corresponding to the initial and final centers' states $|\mu\rangle$ and $|\nu\rangle$, respectively. 
\begin{figure}
 \includegraphics[width=0.3\textwidth]{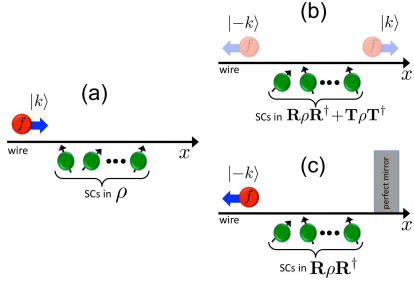}
\caption{(Color online){ {Scheme working principle. (a) $f$ impinges on a set of SCs in $\rho$. (b) After the interaction, $f$ is either reflected or transmitted while the SCs undergo a non-unitary quantum map. (c) With a perfect mirror beyond the centers, $f$ can be only back reflected and the unitary $\hat{R}$ is applied to the SCs.}}  
 \label{Fig1}}
\end{figure} 
Defining a reflection (transmission) operator $\hat{R}$ ($\hat{T}$) in the Hilbert space of the SCs such that $\langle \nu |\hat{R}|\mu\rangle\ug{\bf R}_{\nu\mu}\ug r_{\nu\mu}$ ($\langle \mu |\hat{T}|\nu\rangle\ug{\bf T}_{\nu\mu} \ug t_{\nu\mu}$) \eq(\ref{psif}) can be arranged as $|\Psi_f\rangle\ug|\meno k\rangle \hat{R}|\mu\rangle\piu|k\rangle \hat{T}|\mu\rangle$. Tracing over $f$, the final SCs density operator reads $\hat{R}|\mu\rangle\langle \mu|\hat{R}^\dagger\piu\hat{T}|\mu\rangle\langle \mu|\hat{T}^\dagger$. Thus when the centers are initially in an arbitrary (in general mixed) state $\rho$ (in general mixed) their final state is given by
\begin{equation}\label{psif2}
\rho'=\mathbf{R}\,\rho\mathbf{R}^\dagger+\mathbf{T}\rho\mathbf{T}^\dagger\,\,.
\end{equation}
The normalization condition $\sum_\nu(|r_{\nu\mu}|^2\piu|t_{\nu\mu}|^2)\ug1$ {$\forall \mu$} entails $\hat{R}\hat{R}^\dagger\piu\hat{T}\hat{T}^\dagger\ug\openone$, where $\openone$ is the identity operator of the SCs' Hilbert space. 
We would like the scattering process to implement a multi-qubit gate, which is  {\it unitary}, between the SCs. In general, this is not the case as is evident from \eq(\ref{psif2}) showing that the SCs  undergo instead a quantum {map} \cite{nc} comprising reflection and transmission channels. 

Such hindrance can be overcome in a very natural fashion by simply inserting beyond the centers a {\it perfect mirror} [see  \fig1(c)]. 
As this introduces a hard-wall boundary condition (BC) preventing $f$ from trespassing the right end, this way the transmission channel is in fact fully {\it suppressed}. 
Hence, $t_{\nu\mu}\!\equiv\!0$ and (\ref{psif2}) reduces to
\begin{equation}\label{psif3}
\rho'=\mathbf{{R}}\,\rho\mathbf{{R}}^\dagger\,\,,\nonumber
\end{equation}
where now $\hat{R}\hat{R}^\dagger\ug\hat{R}^\dagger\hat{R}\ug\openone$, \ie {\it in the presence of the mirror $\hat{R}$ becomes a unitary gate}. Thereby, for a monochromatic wavepacket such gate is deterministically implemented whenever $f$ scatters from the centers. Remarkably, this holds regardless of the specific scattering potential. Rather, this affects only the {\it type} of achieved gate. 

Having illustrated how a hard-wall BC guarantees the process unitarity, it is now natural to wonder whether there exist elastic one-channel scattering processes allowing for multi-qubit gates  {\it universal} for QC \cite{nc}. We will show that this is indeed the case. With this aim, we first focus on a simple paradigmatic setting, {\it setup A}, comprising two spin-1/2 centers, \ie two qubits \cite{nc}, each {coupled to} a massive particle embodying $f$. We identify {a regime} such that a CZ gate, universal for QC \cite{nc}, naturally arises. We next address {\it setup B}, {comprising multi-level atom-like systems and photons and within experimental reach} in several scenarios \cite{pc, nws, delft,fibers,wallraf}. We show that this can work as an effective emulator of setup A.
\begin{figure}
 \includegraphics[width=0.35\textwidth]{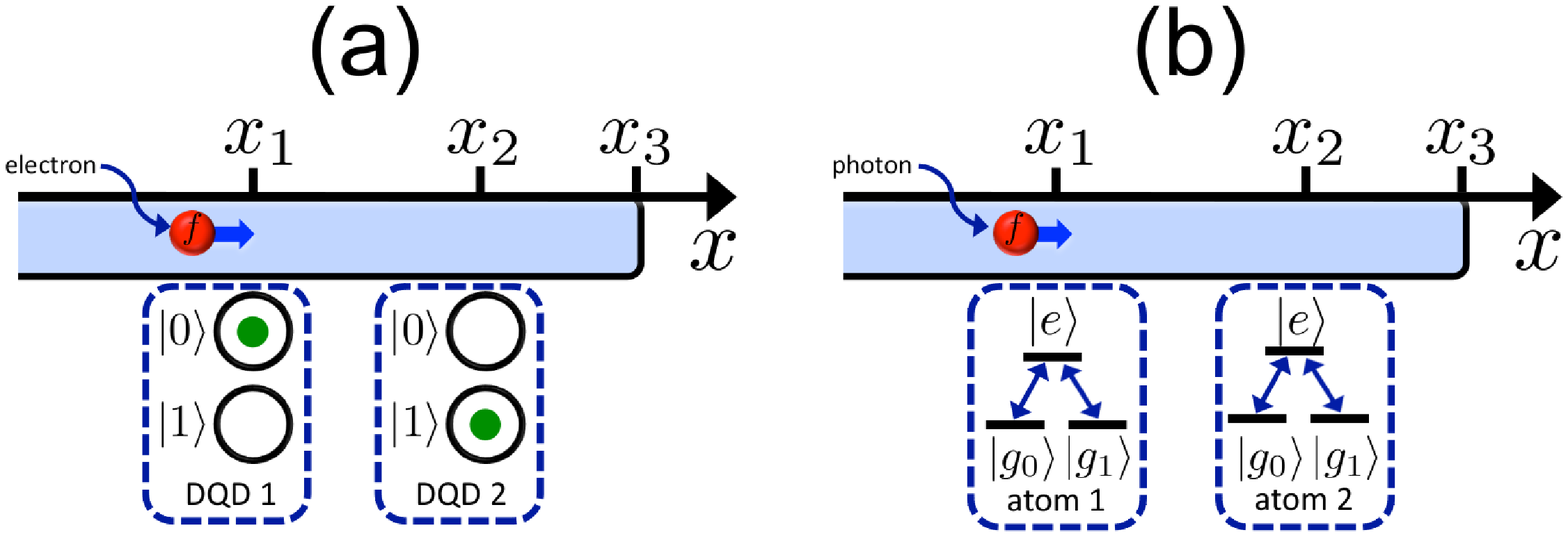}
\caption{(Color online){ Setups implementing our scheme. (a) An electron along a quantum wire with DQDs. (b) A photon along a waveguide with $\Lambda$-type atom-like systems. The right end of the wire/waveguide behaves as a perfect mirror.} 
 \label{Fig1}}
\end{figure} 

{\it Setup A}. This setting [see \fig2(a)] comprises two identical spin-1/2 scattering centers, 1 and 2, lying along a semi-infinite wire along the $x$-axis at $x\!=\!x_1$ and $x\!=\!x_2$, respectively, each coupled to a scattering particle $f$ of mass $m$. The wire ends at $x\!=\!x_3$. Let $\{|0\rangle_i, |1\rangle_i\}$ be an orthonormal basis for the $i$th center ($i\!=\!1,2$). In practice, one can consider a semiconducting quantum wire or a carbon nanotube \cite{wires} where an electron populating the lowest subband can undergo scattering from two double quantum dots (DQDs) \cite{dqd} to which it is electrostatically coupled. As shown in \fig2(a), each single-electron DQD is in $|0\rangle$ ($|1\rangle$) if the upper (lower) dot is occupied, hence implementing an effective qubit \cite{dqd} (tunneling between the upper and lower dots is negligible). Also, the electrostatic {coupling} is negligible for state $|1\rangle$. 
The Hamiltonian is thus modeled as (we set $\hbar\!=\!1$ throughout)
\begin{equation}\label{H}
\hat{H}=\frac{\hat{p}^2}{2m}+\Gamma \delta(x\!-\!x_1)\,|0\rangle_1\langle0|+\Gamma \delta(x\!-\!x_2)\, |0\rangle_2\langle0|\,\,,
\end{equation}
where
$\hat{p}\!=\!-i\, d/dx$ is the momentum operator of $f$, while $\Gamma$ is the height of each contact potential scattering centered at $x\ug x_i$ ($i\ug1,2$){\cite{note-mirror}}. Note that the scattering potential in (\ref{H}) is {\it dispersive} because it cannot induce either 1 or 2 to flip between $|0\rangle$ and $|1\rangle$. {Upon scattering}, each initial SCs' state $|\alpha_1\alpha_2\rangle_{12}$ ($\alpha_i\!=\!0,1$) simply but crucially picks up its own phase shift.
It is trivially checked that for a given state $\mbox{\boldmath$\alpha$}\!\equiv\!\{\alpha_1,\alpha_2\}$ $\hat{H}$ takes the effective form
{$\hat{H}_{\mbox{\boldmath$\alpha$}}\ug\frac{\hat{p}^2}{2m}\piu\sum_{i=1,2}\!\Gamma\delta_{\alpha_i0}\,\delta ({x\!-\!x_i})$.}
The problem thus reduces to a particle scattering from spin-less potentials. We label with $r_{\mbox{\boldmath$\alpha$}}$ 
the $f$'s reflection probability amplitude corresponding to $\hat{H}_{\mbox{\boldmath$\alpha$}}$, where the subscript here specifies both the initial and final centers' state (these coincide owing to the dispersive interaction). Likewise, in the DQDs computational basis $\{|00\rangle,|01\rangle,|10\rangle,|11\rangle\}$ ${\bf R}$ necessarily has the diagonal form $ \mathbf{R}\ug{\rm diag(r_{00},r_{01},r_{10},r_{11})}$.

{Adopting a standard procedure}, to derive $r_{\mbox{\boldmath$\alpha$}}$ we assume that $f$ is left-incoming and seek the stationary state $\Psi_{\mbox{\boldmath$\alpha$}}(x)$ fulfilling $\hat{H}_{{\mbox{\boldmath$\alpha$}}}\Psi_{\mbox{\boldmath$\alpha$}}(x)\ug k^2/(2m)\Psi_{\mbox{\boldmath$\alpha$}}(x)$ of the form $\Psi_{\mbox{\boldmath$\alpha$}}(x)\ug\Psi_{{\mbox{\boldmath$\alpha$}}+}(x)\piu\Psi_{{\mbox{\boldmath$\alpha$}}-}(x)$ with (for simplicity we drop the dependance on ${\mbox{\boldmath$\alpha$}}$ whenever unnecessary)
\begin{eqnarray}\label{statstateR}
\Psi_{+}(x)&\ug&\frac{1}{\sqrt{2\pi}}\! \left\{  \theta[(-x)\piu a_1\! \left[ \theta(x)\meno \theta(x\meno x_2)\right]\piu a_2 \theta(x\meno x_2)   \right\}e^{i k x}\,\,,\\
\Psi_{-}(x)&\ug& \frac{1}{\sqrt{2\pi}}\!\left\{ r\,\theta(-x)\piu b_1\! \left[ \theta(x)\meno \theta(x\meno x_2)\right]\piu b_2 \theta(x\meno x_2)   \right\}\!e^{-i k x},\,\,\,\,\,\,\,\label{statstateL}
\end{eqnarray} 
where $k\!>\!0$, $\theta(x)$ is the Heaviside step function and we set $x_1\ug0$. Due to \eqs(\ref{statstateR}) and (\ref{statstateL}), $\Psi_{+}(x)$ [$\Psi_{-}(x)$] represents the right-propagating (left-propagating) part of  $\Psi(x)$. Note that $\Psi(x)$ is specified by the five ${{\mbox{\boldmath$\alpha$}}}$-dependent coefficients $\{r ,a_1,b_1,a_2,b_2\}$. These are found by requiring that $\Psi(x)$ and its derivative with respect to $x$ $\Psi'(x)$ match the five BCs
\begin{eqnarray}\label{5bcs}
\Psi(x_i^-)&=&\Psi(x_i^+)\,\,\,\,\,\,(i\ug1,2)\,,\,\,\,\,\,\,\,\,\,\,\,\Psi(x_3)=0,\label{bc123}\\
\Delta  \Psi'|_{x_i}&=&2m\,\Gamma\delta_{\alpha_i0}\,\Psi(x_i)\,\,\,\,\,\,(i\ug1,2)\,\,.\label{bc45}
\end{eqnarray}
{\eqs(\ref{bc123}) ensure the matching of the wave function at the centers' locations (first two) and the hard-wall BC owing to the end of the wire at $x\!=\!x_3$ (latter equation)}. \eqs(\ref{bc45}) are standard constraints on the discontinuity of $\Psi'(x)$ at the centers' positions $\Delta \Psi'|_{x_i}\ug \Psi'(x_i^+)\meno\Psi'(x_i^-)$ {due to} the $\delta$-potentials in (\ref{H}) {[they are derived by integrating the Schr\"odinger equation (SE) corresponding to $\hat{H}_{\mbox{\boldmath$\alpha$}}$ over an infinitesimal range across $x\ug x_1$ and $x\ug x_2$].} 
{Solving the linear system (\ref{bc123})-(\ref{bc45})} we end up with 
\begin{eqnarray}\label{rs}
r_{{\mbox{\boldmath$\alpha$}}}&\ug& -\exp\!\left\{2i\arg\left[{\exp(-i  kx_{31}})\meno 
  2 \gamma \delta_{\alpha_20} [\cos{k x_{21}} \right.\right.\nonumber\\
  &&\,\,\,\,\,\,\,\,\,- (i\piu 2\gamma \delta_{\alpha_10}\!) \sin{kx_{21}}] \!\sin k x_{32}
  \left.\left.\meno
  2 \gamma \delta_{\alpha_10} \!\sin k x_{31}\right]\right\},\,\,\,\,\,\,\,\,
\end{eqnarray}
where $x_{ij}\ug x_{i}\meno x_j$ and $\gamma\ug m\Gamma/k.$
As expected, this yields $|r_{\mbox{\boldmath$\alpha$}}|\ug1$ regardless of all the parameters and ${\mbox{\boldmath$\alpha$}}$, namely $f$ is reflected back with certainty. 
Consider the regime 
\begin{eqnarray}\label{regime}
kx_{21}=n\pi,\,\,\,\,\,\,\,kx_{32}=(n'\piu1/2)\,\pi\,\,\,\,\,\,\,\gamma\gg1\,\,,
\end{eqnarray}
where $n$ and $n'$ are arbitrary integers.
{Replacing} (\ref{regime}) in (\ref{rs}) and using that $k x_{31}\ug kx_{32}\piu kx_{21}$, we end up with
\begin{equation}\label{gate}
r_{00}=r_{01}=r_{10}=-r_{11}=  -1\,\,,
\end{equation}
which yields the gate matrix ${\bf R}\ug {\rm diag}(1,1,1,-1)$ (up to an irrelevant global phase factor),
\ie the well-known CZ gate \cite{nc}. This proves that a single scattering process can implement a {universal TQG}. {Intuitively}, unlike \rref\cite{guillermo} here the geometry secures unitarity leaving the physical parameters free to be tuned so as to match a CZ.

{\it Setup B}. {In this setting [see \fig2(b)] $f$ {is} a photon propagating along a 1D waveguide, having geometry analogous to setup A, and scattering from two three-level atom-like systems}. Each ``atom" $i\ug1,2$ has a $\Lambda$-type energy-level configuration consisting of a twofold-degenerate ground doublet spanned by states $\{|{g_0}\rangle, |{g_1}\rangle\}$ and an excited state $|e\rangle$ [see \fig2(b)]. Hence, a two-photon Raman transition between $|g_0\rangle$ and $|g_1\rangle$ can occur through absorption and re-emission of a scattering photon. Assuming a linear photon dispersion law $\omega\ug \upsilon k$ with $\omega$ the photon energy and $\upsilon$ the group velocity, the free-field Hamiltonian in the waveguide reads~\cite{shen_fan,sorensen}
$\hat{H}_{f}\!=\!-i\sum_{d\!=\!\pm}\int dx\,
\upsilon_{d}\,\hat{c}_{d}^\dag(x){\partial}_x\hat{c}_{d}(x)$ with $\upsilon_{+}\!=\!-\upsilon_{-}=\upsilon$ and
$\hat{c}_{+}^\dag(x)$ [$\hat{c}_{-}^\dag(x)$] the
bosonic operator creating a right (left) propagating photon at $x$. The free atomic Hamiltonian reads $\hat{H}_{a}\ug\omega_0\sum_{i=1,2}  |e\rangle_i\langle e|$, where $\omega_0$ is the energy gap between $|e\rangle$ and the ground doublet.
The field-$i$th atom coupling is modeled as \cite{sorensen} ${\hat{H}_{fi}}\ug J \int dx \,\delta(x\meno x_i)\,[\hat{c}(x)\hat{S}_i^\dagger\piu{\rm H.c.}]$ (under the usual rotating-wave approximation), where $\hat{c}(x)\ug \hat{c}_{+}(x)\piu \hat{c}_{-}(x)$ annihilates a photon at $x$ regardless of its propagation direction, $J$ is the rate associated with each transition $|g_j\rangle_i\!\leftrightarrow\!|e\rangle_i$ ($\forall j\ug0,1$) and $\hat{S}_i^\dagger\ug\sum_j |{e}\rangle_i\langle{g_j}|$. The full Hamiltonian thus reads $\hat{H}\ug\hat{H}_f\piu\hat{H}_a\piu \hat{H}_{f1}\piu\hat{H}_{f2}$. It is convenient to use as basis of each ground doublet the symmetric and antisymmetric combinations of the two ground states $|\phi^\pm\rangle_i\ug(|g_0\rangle_i\pm|{g_1}\rangle_i)/\!\sqrt{2}$. As $|\phi^-\rangle_i$ is a dark state \cite{sorensen}, \ie $\hat{S}_i^\dagger|\phi^-\rangle_i\ug0$, the atomic raising operator takes the effective form $\hat{S}_i^\dagger\!\equiv\! \sqrt{2}|{e}\rangle_i\langle{\phi^+}|$. Thus the {Raman process} does not couple $|{\phi^+}\rangle$ and $|{\phi^-}\rangle$. It should be clear now that by taking $|{0}\rangle\ug|{\phi^+}\rangle$ and $|{1}\rangle\ug|{\phi^-}\rangle$ for each atom, as long as these are initially in the ground doublet, setup B in fact possesses all the {key} features of A. Indeed, if the $i$th atom is in $|{1}\rangle_i\ug|{\phi^-}\rangle_i$ the corresponding potential $\hat{H}_{fi}$ vanishes. If in $|{0}\rangle_i$, {it may undergo a second-order transition $|{0}\rangle_i\!\rightarrow\!|{e}\rangle_i\!\rightarrow\!|{0}\rangle_i$ so as to eventually pick up a phase shift once $f$ is scattered off}. To make rigorous such considerations, we {next prove} that the reflection coefficients are again, with due replacements, given by (\ref{rs}) as for setup A.

$\hat{H}$ conserves the total number of excitations. Thus, following a standard approach \cite{shen_fan,sorensen}, we seek one-excitation stationary states of the form 
\begin{eqnarray}\label{statstate2}
|\Psi_{\mbox{\boldmath$\alpha$}}\rangle\ug\sum_{d=\pm}\!\int\!dx\,\psi_{{\mbox{\boldmath$\alpha$}}d}(x)\hat{c}_{d}^\dagger(x)|{ {\rm vac}}\rangle|\alpha_1\alpha_2\rangle_{12}\nonumber\\
+ \varepsilon_1|{ {\rm vac}}\rangle|{e}\alpha_2\rangle_{12}\piu\varepsilon_2|{ {\rm vac}}\rangle|{\alpha_1}e\rangle_{12},
\end{eqnarray}
where $\psi_{{\mbox{\boldmath$\alpha$}}\pm}(x)$ have a form analogous to \eqs(\ref{statstateR}) and (\ref{statstateL}) thus being specified by parameters $\{r,a_1,a_2,b_1,b_2\}$, $\{\varepsilon_i\}$ are excited-state amplitudes and $|{\rm vac}\rangle$ is the field vacuum state. Thus, for given ${\mbox{\boldmath$\alpha$}}$  $|\Psi_{\mbox{\boldmath$\alpha$}}\rangle$ is specified by the 7 complex amplitudes $\{r,a_1,a_2,b_1,b_2,\varepsilon_1,\varepsilon_2\}$ {and} obeys the SE $\hat{H}|\Psi_{\mbox{\boldmath$\alpha$}}\rangle\ug \upsilon k |\Psi_{\mbox{\boldmath$\alpha$}}\rangle$. Projecting this onto $c_{\pm}^\dagger(x)|{\rm vac}\rangle|{\alpha_1\alpha_2}\rangle$ gives (we henceforth omit subscript ${\mbox{\boldmath$\alpha$}}$)
\begin{eqnarray}\label{eq1}
\mp i \upsilon\psi_{\pm}'(x)\piu \tilde{J} \sum_{i=1,2} \delta_{\alpha_i0} \varepsilon_{i}\delta(x\meno x_i)\ug\upsilon k\, \psi_{\pm}(x)\,\,,\,\,\,\,\,\,\,\,\,
\end{eqnarray}
where $\tilde{J}\ug\sqrt{2}J$ is the rate associated with the transition $|{0}\rangle\!\leftrightarrow\!|{e}\rangle$.
Further projection of the SE onto $|{\rm vac}\rangle|{e\alpha_2}\rangle$ and  $|{\rm vac}\rangle|{\alpha_1e}\rangle$ immediately yields that for each $i\ug1,2$ $\varepsilon_i\ug \tilde{J} \delta_{\alpha_i0}\psi(x_i)/(\upsilon k\meno \omega_0)$. Replacing these in (\ref{eq1}) we are left with  $\{\psi(x),\psi_\pm(x)\}$ only
\begin{eqnarray}\label{eq2}
\mp i \upsilon\psi_{\pm}'(x)\piu \tilde{J}\,^2/(\upsilon k\meno\omega_0) \!\sum_{\ell=1,2}\delta_{\alpha_\ell0}\psi(x_\ell)\delta(x\meno x_\ell)\ug\upsilon k\, \psi_{\pm}(x)\,.\,\,\,\,\,\,\,\,\,\,\,
\end{eqnarray}
Subtracting now the equation for $\psi_-$ from the  $\psi_+$'s one yields $-i \psi'(x)\ug k[\psi_+(x)\meno\psi_-(x)]$, which trivially entails that $-i \Delta\psi'|_{x_\ell}\ug k(\Delta \psi_+|_{x_\ell}\meno\Delta \psi_-|_{x_\ell})$ holds as well for each $\ell\ug1,2$. Each $\Delta \psi_{\pm}|_{x_\ell}$ on the \rhs of the above can be evaluated by integrating (\ref{eq2}) over an infinitesimal interval across $x\ug x_\ell$ ($\ell\ug1,2$), which straightforwardly yields $\Delta \psi_{\pm}|_{x_\ell}\ug \mp(i/\upsilon)\tilde{J}\,^2/(\upsilon k\meno\omega_0) \delta_{\alpha_\ell0}\psi(x_\ell)$. Thereby
\begin{equation}\label{bc45bis}
\Delta  \psi'|_{x_\ell}=2\,\frac{k}{\upsilon}\,\frac{\tilde{J}\,^2}{\upsilon k\meno\omega_0}\,\delta_{\alpha_\ell0}\,\psi(x_\ell)\,\,\,\,\,\,(\ell\ug1,2)\,\,.
\end{equation}
{This} is identical to \eq(\ref{bc45}) once we set
\begin{equation}\label{Gammaeff}
\Gamma\ug \frac{\tilde{J}\,^2}{\upsilon k\meno\omega_0}
\end{equation}
and note that, due to the parabolic dispersion law, in setup A $m\ug k/\upsilon$.

As $\psi(x)$, besides (\ref{bc45bis}), must fulfill conditions analogous to \eqs(\ref{bc123}) {due to} the common geometry of setups A and B, we conclude that amplitudes $\{r_{\mbox{\boldmath$\alpha$}}\}$ for setting B are {\it identical} to (\ref{rs}) with the effective mass and potential height given by $k/\upsilon$ and  (\ref{Gammaeff}), respectively. {In passing, note that $|r_{{\mbox{\boldmath$\alpha$}}}|^2\ug1$ showing that if $f$ is absorbed it will be re-emitted with certainty (each atom behaves as an effective qubit encoded in $\{|g_0\rangle,|g_1\rangle\}$)}.
It {is now}  evident that setup B can be used as an emulator of A, thereby allowing for occurrence of the CZ gate [\cf \eq(\ref{gate})] under conditions (\ref{regime}). Interestingly, the requirement $\gamma\!=\!\Gamma/k\!\gg\!1$ now in fact becomes the {\it resonance condition} (RC) $\upsilon k\!\simeq\!\omega_0$. This agrees with \cite{shen_fan}, where it was shown that the reflectivity of an atomic scatterer becomes unitary in this limit. {Also, although of a different nature the first two requirements in (\ref{regime}) are RCs either: The CZ gate thus stems from a combination of RCs}. 
As anticipated, setup B can be experimentally implemented in several different ways, including photonic-crystal waveguides with defect cavities \cite{pc}, semiconducting (diamond) nanowires with embedded QDs (nitrogen vacancies) \cite{nws,delft}, optical or hollow-core fibers interacting with atoms  \cite{fibers} and microwave (MW) transmission lines coupled to superconducting qubits \cite{wallraf}. {In particular, highly efficient coupling between the dot and the fundamental waveguide mode has been recently achieved in tapered InP nanowires with embedded InAsP QDs \cite{delft}. Interestingly, it was recently shown \cite{solano} that a hard-wall BC  analogous to ours [\cf \fig2(b)] can benefit MW photodetection. {It is also worth mentioning that some features of our scheme are reminiscent of \rrefs \cite{kimble} and \cite{guo}.}

{\it Gate working condition}. In practice, the incoming $f$ has a narrow but finite uncertainty $\Delta k$ around a carrier wave vector $k_0$ fulfilling (\ref{regime}). As ${\bf R}$ is $k$-dependent, we assessed its resilience against deviations from $k_0$ by using the process fidelity $F$ as a figure of merit \cite{epaps}. In a representative case, $\Delta k\!\lesssim\!5\%k_0$ yields $F\!\gtrsim\!95\%$. Also, $\Delta k$ affects the gate duration $\Delta \tau$ according to $\Delta \tau\!\sim\!1/(\upsilon \Delta k)$ \cite{epaps}, which entails a minimum time $\Delta \tau_{\rm min}\!\sim\!10\,(\upsilon k_0)^{-1}$. If $T_{\rm d}$ is the system's decoherence time the gate works reliably when $\Delta \tau_{\rm min}\!\ll\!T_{\rm d}$, hence the working condition reads $T_{\rm d} \!\gg\!10\,(\upsilon k_0)^{-1}$. This is matched in realistic instances \cite{epaps}.

{\it Conclusions}. {We have shown a strategy to quasi-deterministically carry out multi-qubit gates between static qubits through single flying buses scattering from them. This is effective without demanding post-selection of any kind or iteration. The possibility to naturally implement a universal CZ gate has been proven in two different setups including 1D photonic waveguides coupled to atom-like qubits}. We believe this work can set a significant milestone for future advancements in the area of distributed QIP as well as in the emerging field of quantum optics in 1D waveguides. { The design of a full quantum computing architecture, implementing single-qubit operations besides the proposed multi-qubit gates, will be the subject of a future comprehensive work \cite{inprep}}.

{\it Acknowledgements}. {We acknowledge support from FIRB IDEAS (project RBID08B3FM), the National Research Foundation \& Ministry of Education of Singapore, Leverhulme Trust, the EPSRC, the Royal Society and the Wolfson Foundation.}

\section*{Supplementary material}

{The aim of this Supplementary Material is to carry out a detailed analysis of our scheme's performances by relaxing the assumption (made in the main text) to deal with a monochromatic incoming wavepacket. We first quantify the scheme's resilience to deviations of the wave vector from the optimal value yielding a CZ gate. Specifically, we evaluate how different is the implemented gate from the ideal CZ by using a monochromatic wave packet whose associated wave vector is arbitrarily chosen. This can also be regarded as the fidelity corresponding to a generic harmonic secondary wave when a finite-width wavepacket is sent. Next, under the latter (realistic) hypothesis of dealing with a wavepacket, we work out the characteristic time taken by the scattering process. This turns out to be a function of the group velocity and wavepacket width. Finally, we use jointly the above findings in order to define the gate's working condition in terms of the system's decoherence time.}

\subsection{Gate fidelity}

Here, we analyze the resilience of the two-qubit CZ gate proposed in the main text (setups A and B) to an imperfect setting of $k$, \ie the wave vector of the flying particle. For set $x_{21}$, $x_{32}$, $n$ and $n'$ and in the limit $\gamma\!\gg\!1$, the optimal wave vector $k_0$ fulfills [\cfr\eq(9) in the main text]
\begin{eqnarray}\label{regime}
k_0 x_{21}=n\pi,\,\,\,\,\,\,\,k_0x_{32}=(n'\piu1/2)\,\pi\,\,.
\end{eqnarray}
The corresponding reflection operator $\hat{R}_{k_0}$ coincides with the ideal CZ gate and can be written as
\begin{equation}\label{rk0}
\hat{R}_{k_0}\ug|0\rangle_1\!\langle0|\,\openone_2+|1\rangle_1\!\langle1|\,\hat{Z}_2\,\,
\end{equation}
where $\openone_i$, $\hat{X}_i $, $\hat{Y}_i $ and $\hat{Z}_i$ are respectively the identity operator  and the usual Pauli operators of the $i$th static qubit. Using \eq(8) in the main text in the limit $\gamma\gg1$, the gate corresponding to an arbitrary $k$ is given by
\begin{equation}\label{rk}
\hat{R}_{k}\ug|0\rangle_1\!\langle0|\,\openone_2+|1\rangle_1\!\langle1|\,\left(e^{2i k x_{2}}|0\rangle_2\!\langle0|\piu e^{2i k x_{3}}|1\rangle_2\!\langle1|\right)\,\,,
\end{equation}
which in general does not coincide with (\ref{rk0}). 
To quantify how close is $\hat{R}_k$ to $\hat{R}_{k_0}$, we use the process fidelity $F$ \cite{gatefid} as a figure of merit. This is defined as $F\ug{\rm Tr}(\chi_{0}\chi)$, where the $16\!\times\!16$ matrix $\chi$ fully specifies $\hat{R}_{k}$ according to
\begin{equation}\label{chi}
\hat{R}_{k}\,\rho\, \hat{R}_{k}^\dagger\ug\sum_{m n}\chi_{mn}\,\hat{A}_m \,\rho\, \hat{A}_n^\dagger\,\,,
\end{equation}
where $\rho$ is a generic two-qubit state while $\{\hat{A}_m \}$ is a basis for operators acting on $\rho$ [$\chi_0$ specifies $\hat{R}_{k_0}$ in full analogy with (\ref{chi})]. We choose each $\hat{A}_m$ to be the tensor product between two operators taken from the set $\{\openone, \hat{X}, \hat{Y},\hat{Z}\}$. 
Hence, through the replacements $|0\rangle\!\langle0|\ug(\openone\piu\hat{Z})/2$ and $|1\rangle\!\langle1|\ug(\openone\meno\hat{Z})/2$ in \eqs(\ref{rk0}) and (\ref{rk}) one can exactly derive both $\chi_0$ and $\chi_k$. Upon trace of their matrix product, the gate fidelity can be worked out in the compact analytical form
\begin{equation}\label{F}
F\ug\frac{3\piu2\cos(2k x_2)\meno\cos(2k x_{32})\meno2 \cos(2 k x_3)}{8}\,\,.
\end{equation}

In order to assess the gate resilience, it is convenient to study $F$ as given in (\ref{F}) as a function of the dimensionless parameter $k/k_0$, where $k_0$ fulfills (\ref{regime}). Thereby, $F\ug1$ for $k/k_0\ug1$. In \fig1(a), we set $n\ug1$ and $n'\ug0$ [\cfr\eq(\ref{regime})] and plot $F$ against $k/k_0$. For $|k/k_0| \!\lesssim \!0.4$, $F$ decays as the discrepancy between $k$ and $k_0$ grows (regardless of the sign). Importantly, as can be seen from \fig1(b), relative deviations of $k$ from the optimal value as large as $\sim$5\% still ensure the fidelity to stay above 95\%, which amounts to a quasi-deterministic gate operation.
\begin{figure}
 \includegraphics[width=0.4\textwidth]{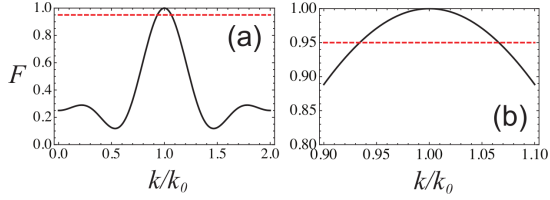}
\caption{(Color online) (a) Gate fidelity $F$ against $k/k_0$. (b) Zoomed view of panel (a) showing that a discrepancy between $k$ and $k_0$ up to over 5\% can be tolerated in that $F$ stays above the 95\% threshold (red dashed line in the plots). 
 \label{Fig1}}
\end{figure} 

\subsection{Gate characteristic time}

Here, we derive the characteristic time $\Delta \tau$ taken by the CZ gate proposed in the main text (setups A and B) in order to be implemented through a single scattering process. To this aim, we will first relax the assumption to deal with a perfectly monochromatic incoming wavepacket for the flying particle (as such, this would clearly entail an infinite $\Delta \tau$). We instead consider a narrow Gaussian wave packet impinging on the static qubits and work out the associated energy-time uncertainty principle $\Delta E$, which is related to the characteristic time by the energy-time uncertainty principle $\Delta E \Delta \tau\! \sim \!1$ (we recall that we set $\hbar\ug1$). 

As incoming flying-particle  wave function $\varphi(x)$ we take the right-propagating Gaussian function
\begin{equation}\label{gaussian}
\varphi(x)\ug\frac{1}{\sqrt{2\pi}\Delta k}e^{-\frac{(x- x_0)^2}{2 \Delta k^2}}e^{i k_0 x}
\end{equation}
specified by the center $x_0$, the carrier wave vector $k_0\!>\!0$ and the wave-vector width $\Delta k$. We assume that $x_0$ and $\Delta k$ fulfill 
\begin{equation}\label{cond-kin}
|x_1\meno x_0|\ge 3\Delta x\,\,,\,\,\,\,\,\,k_0\!\ge\!3\Delta k\,\,,
\end{equation}
where $\Delta x$ is the wave-packet width in position space.
The first of the above conditions specifies that the incoming wave packet is initially fully out of the scattering region (we recall that this lies at $x_1\!\le\!x\!\le\! x_3$, see \fig2 in the main text). The second condition sets an upper bound on the wave-vector width ($\sim\!0.3\,k_0$) so as to ensure that the incoming wave packet $\varphi(x)$ comprises only positive-$k$  secondary plane waves (\ie right-propagating ones). Owing to the latter feature the $\varphi(x)$'s Fourier transform $\tilde{\varphi}(k)\ug1/\!\sqrt{2\pi}\int_{-\infty}^{\infty} dx\, e^{-i k x}\varphi(x)$ thus obeys
\begin{equation}\label{cond-phik}
\tilde{\varphi}(k\le0)\simeq 0\,\,.
\end{equation}
We will next show that the above features of the incoming wave packet guarantee that the characteristic time $\Delta \tau$ is the same as the one associated with the free-propagating particle, \ie in the absence of the static qubits and mirror. The physical mechanism behind this behavior is essentially the one in Ref.~\cite{time-evolution}, where however a different Hamiltonian model was addressed. 
The main features of the demonstration are basically common to both setup A and B. We will carry out the demonstration in detail for the latter since this is somewhat more involved owing to the second-quantization formalism of the associated model.

\subsection{Initial state}

If the static qubits are in the state $|\mbox{\boldmath$\alpha$}\rangle_{12}$ (see main text) and the incoming flying particle is described by (\ref{gaussian}), the initial system's state reads
\begin{equation}\label{psi0}
|\Psi_0\rangle\ug\int_{-\infty}^{\infty}dx\,\varphi(x) \hat{c}^\dagger(x)|{\rm vac}\rangle|\mbox{\boldmath$\alpha$}\rangle_{12}\,\,.
\end{equation}
We first show that due to (\ref{cond-phik}) in (\ref{psi0}) $\hat{c}^\dagger(x)$ can be replaced with $\hat{c}_+^\dagger(x)$ (bosonic operator creating a right-propagating photon at $x$, see main text). As $\hat{c}^\dagger(x)\ug\hat{c}_+^\dagger(x)\piu\hat{c}_-^\dagger(x)$ we need to prove that $\int_{-\infty}^{\infty}dx\,\varphi(x) \hat{c}_-^\dagger(x)|{\rm vac}\rangle\!\simeq\!0$. By replacing in the latter expression the Fourier decomposition $\hat{c}_-(x)\ug\int_{-\infty}^0 dk\,e^{i k x} \hat{c}_k$ ($\hat{c}_k$ is a standard photonic annihilation operator in the $k$-space) we end up with
\begin{eqnarray}\label{pro}
\int_{-\infty}^{\infty}dx\,\varphi(x) \hat{c}_-^\dagger(x)|{\rm vac}\rangle&\ug& \int_{-\infty}^0\!dk\,\frac{\int_{-\infty}^\infty dx\,e^{-i k x} \varphi(x)}{\sqrt{2\pi}} c_k^\dagger\,|{\rm vac}\rangle\nonumber\\
&\ug& \int_{-\infty}^0\!dk\, \tilde{\varphi}(k)\,c_k^\dagger\,|{\rm vac}\rangle\,\!\simeq\!0\,\,,
\end{eqnarray}
where in the last step we made use of (\ref{cond-phik}). Thereby, the initial state is well approximated as
\begin{equation}\label{psi0app}
|\Psi_0\rangle\!\simeq\!\int_{-\infty}^{\infty}dx\,\varphi(x) \hat{c}_+^\dagger(x)|{\rm vac}\rangle|\mbox{\boldmath$\alpha$}\rangle_{12}\,\,.
\end{equation}

\subsection{Spectral decomposition}

To eventually calculate $\Delta E$ we need to derive the spectral decomposition of $|\Psi_0\rangle$ with respect to the system's stationary states $|\Psi_{k\mbox{\boldmath$\alpha$}}\rangle$ (it is now convenient to explicitly indicate the dependance of these on $k$). Use of \eq(11) in the main text yields
\begin{equation}\label{overlap1}
\langle\Psi_{k\mbox{\boldmath$\alpha$}}|\Psi_0\rangle\ug\sum_{d=\pm}\int_{-\infty}^{\infty}dx\,dx'\psi_{k\mbox{\boldmath$\alpha$}d}^*(x)\varphi(x)\langle {\rm vac}|\hat{c}_d(x)\hat{c}_+^\dagger(x')|\rm{vac}\rangle\,\,\,.
\end{equation}
Next, using that the vacuum state vanishes whenever any annihilation operator is applied to it we obtain $\langle {\rm vac}|\hat{c}_d(x)\hat{c}_+^\dagger(x')|\rm{vac}\rangle\ug$$\ug[\hat{c}_d(x),\hat{c}_+^\dagger(x')]$. Such commutator is zero for $d\ug-$ because any right-propagating bosonic operator commutes with any left-propagating one. We thus need to calculate only $[\hat{c}_+(x),\hat{c}_+^\dagger(x')]$. To this aim, we use  $\hat{c}_+(x)\ug\int_{0}^\infty dk\,e^{i k x} \hat{c}_k$ and $[\hat{c}_k,\hat{c}_{k'}]\ug\delta(k\meno k')$ and obtain
\begin{eqnarray}\label{comm}
[\hat{c}_+(x),\hat{c}_+^\dagger(x')]&\ug&\delta(x\meno x')-\frac{1}{2\pi}\int_{-\infty}^0dk\,e^{i k (x- x')}\nonumber\\
&\ug&\frac{\delta(x\meno x')}{2}\piu \frac{i}{2\pi(x\meno x')}\,\,.
\end{eqnarray}
Hence, the scalar product in (\ref{overlap1}) becomes
\begin{equation}\label{overlap2}
\langle\Psi_{k\mbox{\boldmath$\alpha$}}|\Psi_0\rangle\ug  \int_{-\infty}^\infty \!dx\,\psi_{k\mbox{\boldmath$\alpha$}+}^*(x)\left(\frac{\varphi(x)}{2}-\frac{i}{2\pi}\int_{-\infty}^\infty dx'\frac{\varphi(x')}{x'\meno x}\right)\,\,.\,\,\,\,\,\,\,\,\,
\end{equation}
The principal value of the improper integral on the \rhs can be worked out in a standard way with the help of the Cauchy integral theorem and formula, the Jordan lemma and (\ref{cond-phik}). Note that owing to the latter only positive-$k$ plane waves appear in the Fourier expansion of $\varphi(x)$ whose extension to the complex plane $\varphi(x\piu i y)$ thus vanishes for $y\!\rightarrow\!\infty$. Hence, the integral is well approximated as $\int_{-\infty}^\infty dx'{\varphi(x')}/{(x'\meno x)}\!\simeq\!i\pi\varphi(x)$ and thereby (\ref{overlap2}) is simply given by 
\begin{equation}\label{overlap3}
\langle\Psi_{k\mbox{\boldmath$\alpha$}}|\Psi_0\rangle\!\simeq\!  \int_{-\infty}^\infty \!dx\,\psi_{k\mbox{\boldmath$\alpha$}+}^*(x)\varphi(x)\,\,.\,\,\,\,\,\,\,\,\,
\end{equation}
To evaluate the \rhs of the above equation, we use \eq(4) in the main text along with the first of (\ref{cond-kin}). Owing to the latter, $\varphi(x)$ is negligible for $x\!\ge\!x_1$. It should now be clear that by setting $x_1\ug0$ the contributions to the integral proportional to $\theta(x)\meno\theta(x\meno x_2)$ and $\theta(x\meno x_2)$ [\cfr \eq(4)] can be neglected while the first Heaviside step function $\theta(\meno x)$ can be omitted for all practical purposes. Thereby, we are finally left with
\begin{equation}\label{overlap4}
\langle\Psi_{k\mbox{\boldmath$\alpha$}}|\Psi_0\rangle\!\simeq\!  \frac{1}{\sqrt{2\pi}}\int_{-\infty}^\infty \!dx \,e^{-i k x}\varphi(x)\equiv \tilde{\varphi}(k)\,\,.\,\,\,\,\,\,\,\,\,
\end{equation}
\subsection{Characteristic time}
By recalling that the energy of $|\Psi_{k\mbox{\boldmath$\alpha$}}\rangle$ is $E\ug \upsilon k$ the usual decomposition in the steady-state basis of the evolved state at a time $t$ reads
\begin{equation}\label{psit}
|\Psi(t)\rangle\ug\int \!dk\, \langle\Psi_{k\mbox{\boldmath$\alpha$}}|\Psi_0\rangle\,e^{-i \upsilon k t}\,|\Psi_{k\mbox{\boldmath$\alpha$}}\rangle\!\simeq\!\int \!dk\,  \tilde{\varphi}(k)\,e^{-i \upsilon k t}\,|\Psi_{k\mbox{\boldmath$\alpha$}}\rangle\,\,,
\end{equation}
where we have used the result (\ref{overlap4}). \eq(\ref{psit}) is analogous to the case where the  flying particle propagates freely but for the replacement of each plane wave with the steady state of the full scattering problem corresponding to the same $k$. Hence, while the dynamics will be obviously different the energy uncertainty is necessarily given by $\Delta E\ug \upsilon \Delta k$ as in the free-particle case. Using $\Delta E\Delta \tau\!\sim\!1$, it follows that the characteristic time is also the same in the two cases and simply reads
\begin{equation}\label{deltat}
\Delta \tau\!\sim\!\frac{1}{\upsilon\Delta k}\,\,.
\end{equation}
These findings are in line with the content of Ref.~\cite{time-evolution}.

As for setup A (see related section in the main text), the demonstration proceeds analogously but is more straightforward because of the first-quantization picture. Indeed, one assumes an incident wave packet of the form (\ref{gaussian}) fulfilling conditions (\ref{cond-kin}). Then, using these, \eq(\ref{overlap3}) is immediately worked out (without the need for dealing with commutation rules) and next \eq(\ref{overlap4}) is found. \eq(\ref{psit}) still holds aside from the replacement of the linear dispersion relation with $E\ug k^2/(2m)$ (see main text). The corresponding free-particle energy uncertainty now becomes $\Delta E\ug \upsilon_k\Delta k$, where $\upsilon\ug k/m$ is the group velocity corresponding to $k$. Finally, we end up with $\Delta \tau\!\sim\!1/( \upsilon_k\Delta k)$.

\subsection{Gate working condition}

To work out the gate optimal working condition, one needs to take into account \eqs(\ref{F}) and (\ref{deltat}) jointly. As shown in the first section of this Supplementary Material, in a representative case the gate fidelity $F$ stays above 95\% for $\Delta k$ up to $\sim\!5\%k_0$. Using (\ref{deltat}), this immediately entails that the shortest time taken by the scattering process to reliably implement the CZ gate is given by
\begin{equation}\label{min}
\Delta \tau_{\rm min}\!\sim\!\frac{1}{\upsilon \,5\% k_0}\!\sim\!{10}\,(\upsilon k_0)^{-1}\,\,.
\end{equation}
\\
Clearly, in order for the gate to be effective one needs the process to be short enough compared to the decoherence time $T_{\rm d}$ (the latter is the typical time to elapse before the system's coherence is significantly spoiled by environmental interactions). We thereby demand that $\Delta \tau\!\ll\!T_{\rm d}$. Since this must occur under the constraint $\Delta \tau\!>\!\Delta \tau_{\rm min}$ we conclude that the gate can work provided that $\Delta\tau_{\rm min}\!\ll\!T_{\rm d}$, which in the ligtht of (\ref{min}) gives
\begin{equation}\label{working}
T_{d}\gg10\,(\upsilon k_0)^{-1}\,\,.
\end{equation}
As representative numerical instances, let us consider a GaAs nanowire with embedded InAs quantum dots \cite{GaAs}. By estimating the group velocity as $\upsilon\ug c/n_{\rm GaAs}$, where $n_{\rm GaAs}\simeq\!3.4$ is the refractive index of GaAs and the typical carrier wavelength as $\lambda_0\ug2\pi/k_0\!\sim\!900\,$nm the above condition yields $T_{d}\!\gg\!1.6\cdot\!\,10^{-14}$s. This is fulfilled since typical decoherence times of quantum dots of this sort are of the order of ps. Next, we address a diamond nanowire where Nytrogen-vacancy (NV) centers work as atom-like systems \cite{diamond}. In such a setting, the refractive index is $\simeq\!2.4$ while $\lambda_0\!\simeq\!640\,$ nm giving $T_{d}\!\gg\!8\cdot\!10^{-15}$s. Again, this is very well matched in the light of decoherence times of NV centers, which typically exceed tens of ns \cite{NV}.

\end{document}